\documentclass[%
 reprint,
nofootinbib,
 amsmath,amssymb,
 aps,
 prd
]{revtex4-2}

\usepackage[T1]{fontenc}
\usepackage[utf8]{inputenc}
\usepackage{lmodern}          
\usepackage{microtype}        
\usepackage{amsmath,amssymb,mathtools}
\usepackage{graphicx}
\usepackage{xcolor}
\usepackage{hyperref}
\usepackage{orcidlink}        

\usepackage{graphics,epsfig,psfig}
\usepackage{todonotes}
\usepackage[normalem]{ulem}
\usepackage{xcolor}
\usepackage{float}
\usepackage{dcolumn}
\usepackage[style=base, singlelinecheck=off, font=small, labelfont=bf, 
justification=raggedright]{caption}
\usepackage{subfigure}
\usepackage[]{inputenc,amssymb}
\usepackage{natbib}
\usepackage{url}
\usepackage{orcidlink}
\usepackage{amsmath}

\usepackage{xcolor}
\definecolor{softgreen}{RGB}{0,150,0}

\hypersetup{
  colorlinks=true,
  linkcolor=blue,
  citecolor=blue,
  urlcolor=blue
}

\newcommand{\be}{\begin{equation}}
\newcommand{\ee}{\end{equation}}

\begin{document}

\title{The First Upper Bound on the Non-Stationary Gravitational Wave Background and its Implication on the High Redshift Binary Black Hole Merger Rate}

\author{Mohit Raj Sah\,\orcidlink{0009-0005-9881-1788}}
\email{mohit.sah@tifr.res.in}
\author{Suvodip Mukherjee\,\orcidlink{0000-0002-3373-5236}}
\email{suvodip.mukherjee@tifr.res.in}
\affiliation{Department of Astronomy and Astrophysics, Tata Institute of Fundamental Research, Mumbai 400005, India}

\begin{abstract}
The high redshift merger rate and mass distribution of black hole binaries provide a direct probe to distinguish astrophysical black holes (ABHs) and primordial black holes (PBHs), which can be studied using the Stochastic Gravitational-Wave Background (SGWB). The conventional analyses solely based on the power spectrum are limited in constraining the properties of the underlying source population under the assumption of a non-sporadic Gaussian distribution. However, recent studies have shown that SGWB is expected to be sporadic and non-Gaussian in nature, which gives rise to non-zero \textit{spectral correlation} that depends on the high redshift merger rate and mass distribution of the compact objects.   
In this work, we present the first spectral covariance analysis of the SGWB using data from the LIGO--Virgo--KAGRA collaboration during the third and the first part of the fourth observing runs. Our analysis indicates that the current spectral correlation is consistent with non-stationary noise, yielding no detection and providing only upper bounds over the frequency range of 20 Hz to 100 Hz. This upper bound on the spectral correlation translates into a mass-distribution-dependent upper bound on the merger rate of PBHs. This provides a stringent upper bound on the PBH merger rate at high redshift and hence puts constraints on the PBH formation scenarios. In the future, detection of this signal will provide a new avenue to probe the high-redshift black hole population using gravitational waves. 
\end{abstract}

\maketitle

\section{Introduction}

The stochastic gravitational-wave background (SGWB) is a key target for current and future generations of gravitational-wave (GW) detectors \citep{romano2017detection,renzini2022stochastic,abac2025upper,agazie2023nanograv,antoniadis2023second,zic2023parkes,xu2023searching}. It arises from the superposition of a large number of individually unresolvable GW sources \citep{christensen2018stochastic}. The most likely contributors to the SGWB are unresolved inspiraling and merging compact binaries that remain below the detection threshold for individual events. Other astrophysical sources, such as a population of core-collapse supernovae, magnetars, etc, may also contribute, although their expected signal strength is subdominant \citep{buonanno2005stochastic,chowdhury2021stochastic}. In addition to these astrophysical sources, the SGWB can also originate from cosmological processes, such as inflation and phase transitions in the early Universe \citep{damour2005gravitational,siemens2007gravitational,brito2017gravitational,tsukada2019first}.

The searches for the SGWB signal rely on cross-correlating strain data from spatially separated detectors to isolate the SGWB signal from uncorrelated instrumental noise \citep{allen1999detecting,thrane2013sensitivity,thrane2009probing}.
 
Other search techniques include the Bayesian Search (TBS) \citep{smith2018optimal}, Cross-Correlation Intermediate (CCI) search \citep{coyne2016cross}, spectrogram correlated stacking \citep{Dey:2023oui}, and search for intermittent SGWB \citep{lawrence2023stochastic}. Traditional searches for the SGWB primarily focus on measuring the energy-density power spectrum $\Omega_{\rm GW}(f)$, implicitly assuming that the background is stationary and Gaussian. However, the power spectrum alone suffers from significant degeneracies among source-population parameters, motivating the need for additional statistical observables \citep{sah2024non}. 

\begin{figure*}
    \centering    \includegraphics[width=0.8\linewidth]{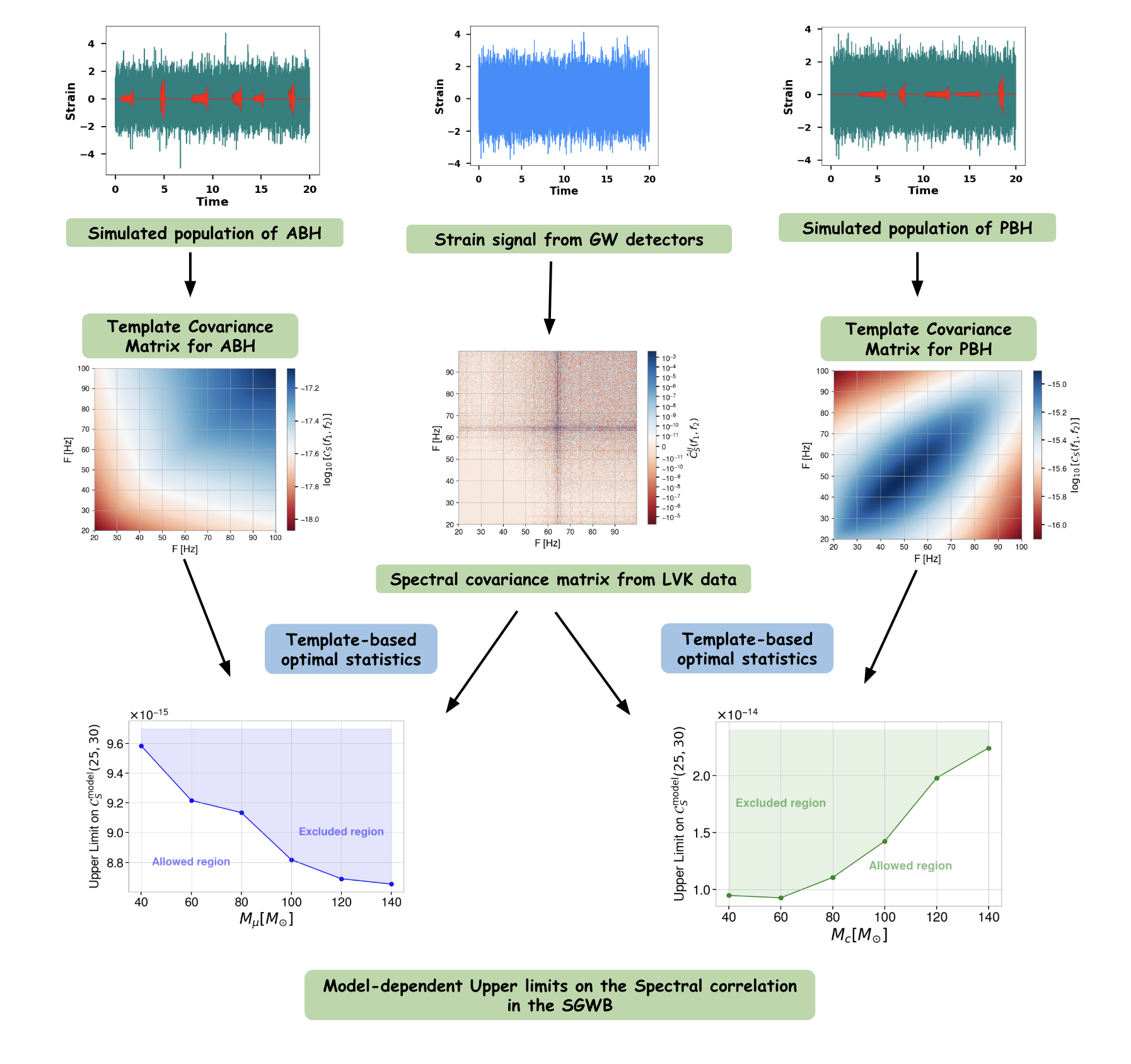}  
    \centering
    \caption{Schematic overview of the analysis pipeline for estimating the spectral covariance in the GW strain data. The procedure begins with the strain data from the LIGO detectors, from which the spectral covariance signal is computed. Model-dependent templates of the spectral covariance matrix are then generated using simulated populations of BHBs. An optimal statistic is constructed by performing a weighted average of the covariance structure with these templates and the noise variance, which is finally used to derive upper limits on the spectral covariance amplitude of the SGWB.}
\label{fig:Motv}    
\end{figure*}

While the assumption of stationarity is valid when the signal is dominated by cosmological processes in the early Universe \citep{romano2017detection,christensen2018stochastic,caprini2018cosmological} or by a very large number of overlapping astrophysical sources, it breaks down for astrophysical SGWBs produced by a finite and discrete population of compact binary coalescences (CBCs). In such cases, Poisson fluctuations in the number of sources contributing within a finite observation time introduce temporal variability in the SGWB amplitude, leading to non-stationary and non-Gaussian features in the signal \citep{mukherjee2020time,sah2024non}. These effects induce correlations between different frequency modes of the SGWB, providing additional observables beyond the power spectrum that encode information about the merger rate evolution and mass distribution of the underlying compact-object population, particularly at high redshift. Building on these insights, recent theoretical work has emphasized that spectral covariance offers a powerful probe of the discrete astrophysical origin of the SGWB \citep{sah2024non}. Fig. \ref{fig:Motv} provides a schematic overview of our analysis pipeline for the search for spectral covariance in GW strain data. We construct the estimator of the SGWB energy density by cross-correlating the strain data from two detectors. From this estimator, we derive the spectral covariance signal in the SGWB density spectrum. We then generate model-dependent templates of the spectral covariance matrix using simulated populations of black hole binaries (BHBs). Finally, we define an optimal statistic by performing a weighted average of the measured spectral covariance structure over all frequency pairs, using these model templates and the corresponding measurement uncertainties. This optimal statistic is then used to place model-dependent upper limits on the spectral covariance amplitude in the SGWB spectrum.

The isotropic SGWB density has been constrained in previous studies \citep{abbott2021upper,abac2025upper}. Similarly, \citet{abbott2021search} investigated anisotropic features using data from the first three observing runs. In this work, we present the first upper limit on the non-stationary nature of the SGWB, derived from publicly available strain data from the O3 and O4a observing runs of LIGO–Hanford (H1) and LIGO–Livingston (L1) \citep{abbott2023open,ligo2025open}. We exclude data from earlier observing runs and from the Virgo detector \citep{abbott2023open,ligo2025open}, as their comparatively higher noise levels would increase computational cost without offering significant improvement in sensitivity.

We place an upper limit on the spectral covariance using a template-weighted optimal estimator, where the templates are constructed from simulated BH populations. The resulting upper limits on the spectral covariance amplitude—represented by the covariance between frequencies of 25 Hz and 30 Hz—lie in the range of $(8.3~-~21.5)\times10^{-15}$ across different population templates. Similarly, we place upper bounds on the merger rate of PBH binaries, $\mathcal{R}_{\rm PBH}(z = 1)$, in the range
($7.2 \times 10^{4}$ - $4.1 \times 10^{2})\,\mathrm{Gpc}^{-3}\,\mathrm{yr}^{-1}$, depending on the characteristic mass scale and the width of the assumed log-normal mass distribution.

This paper is organized as follows. In Sec. \ref{Sec:SGWB}, we discuss the BHB population models along with the SGWB density and its spectral covariance generated by these populations. In Sec. \ref{Meth}, we describe the cross-correlation estimator of the SGWB and the estimator of its spectral covariance. In Sec. \ref{Result}, we present the results of our spectral covariance analysis, and derive upper limits on the amplitude of the spectral correlation of the SGWB density and the merger rate of the high redshift BHB population. Finally, in Sec. \ref{conc}, we summarize our conclusions and discuss prospects for future work.

\section{GW Background from a population of BHB}\label{Sec:SGWB}

\subsection{Population of BHB}

We consider a population of BHBs for both astrophysical black holes (ABHs) and primordial black holes (PBHs). These two populations differ in their merger rate and mass distributions, which in turn imprint distinct signatures on the amplitude and spectral shape of the SGWB and its spectral covariance. In the following subsections, we present the parameterizations adopted for each population.

\subsubsection*{Astrophysical black hole binaries}
We model the coalescence rate of ABH binaries as \cite{abbott2021population,mukherjee2021can,abac2025gwtc}.

\begin{equation}
    \mathcal{R}_{\rm ABH}({\rm z}) \propto \frac{(1+{\rm z})^{\gamma}}{1 + \left(\frac{1+{\rm z}}{1+\rm z_p}\right)^{\kappa}} ,
\end{equation}
where z is the redshift, $\gamma$, $\kappa$, and $\rm z_p$ are the indices characterizing the redshift evolution of the merger rate. The parametric form is motivated by the Madau-Dickinson star formation rate (SFR) \citep{madau2014cosmic}.

We assume a Power-law + Gaussian model for the distribution for the primary mass \citep{abbott2023population,abac2025gwtc}

\begin{equation}
    P_{\rm ABH}(m) = (1-\lambda) {\mathcal{P}}\left(m,\alpha\right) + \lambda ~ \mathcal{G}\left(m, M_{\mu},\sigma_m\right),
\label{Pz1}
\end{equation}
where ${\mathcal{P}}\left(m,\alpha\right)$ and $\mathcal{G}\left(m, M_{\mu},\sigma_m\right)$ are normalized power-law function and Gaussian function, respectively, $\alpha$ is the power-law index, $\lambda$ amplitude of the Gaussian component, and $M_{\mu}$ and $\sigma_m$ are the mean and standard deviation of the Gaussian function, respectively. For the secondary mass, we assume a power-law distribution with the same index $\alpha$. This phenomenological model captures the key features of the BH mass spectrum inferred from GW observations, consistent with expectations from stellar evolution.

\subsubsection*{Primordial black hole binaries}

We also consider a cosmological population of PBH binaries. We model the merger rate of the PBH as 
\begin{equation}
    \mathcal{R}_{\rm PBH}({\rm z}) \propto (1+{\rm z})^{\beta},
    \label{eq:R_PBH}
\end{equation}
where $\beta$ is the index of redshift evolution of merger rate. For most PBH formation scenarios with the Poisson clustering (no clustering), the parameter takes a typical value of $\beta \simeq 1.3$ \citep{carr2017primordial,raidal2017gravitational, sasaki2018primordial,raidal2019formation}. We adopt this as the fiducial value for our analysis. The merger rate of PBHs can be modeled depending on whether the binaries are dominated by Poisson statistics or whether there is clustering. In highly clustered regimes ($\xi_{\rm PBH} \gg 1$), the local merger rate can be exponentially suppressed. The local merger rate in this extreme clustering limit can be expressed as

\begin{equation}
    \begin{aligned}
    \mathcal{R}_{\mathrm{\rm PBH}}({\rm z}=0) \propto ~& 
    \xi_{\mathrm{\rm PBH}}^{0.7}\, f_{\mathrm{\rm PBH}}^{1.7}\,
    \exp\left[-\left(\frac{\xi_{\mathrm{\rm PBH}} f_{\mathrm{\rm PBH}}}{10^{4}}\right)\right],\\
    &\text{for } \xi_{\mathrm{\rm PBH}} f_{\mathrm{\rm PBH}} > 10^{3},
    \end{aligned}
    \label{eq:R_PBH_clustered}
\end{equation}
and for an unclustered (Poisson) distribution of PBHs, the merger rate can be expressed as \citep{raidal2019formation,clesse2022gw190425}
\begin{equation}
    \mathcal{R}_{\mathrm{\rm PBH}}({\rm z}=0)
    \propto  f_{\mathrm{sup}}\, f_{\mathrm{\rm PBH}}^{53/37}\, 
    \eta^{-34/37} \left(m_1 + m_2 \right)^{-32/37},
\label{eq:R_PBH_poisson}
\end{equation}
where $m_1$ and $m_2$ are the component masses of the binary, $f_{\rm PBH}$ is the fraction of PBH in dark matter, $\xi_{\rm PBH}$ quantifies spatial clustering, $f_{\mathrm{sup}}$ is a suppression factor that accounts for the influence of the surrounding matter distribution and interactions with other PBHs, and $\eta = \Big(\frac{m_1 ~m_2}{m_1 + m_2}\Big)^{2}$.\\

The mass distribution of PBHs is modeled as a log-normal distribution \citep{dolgov1993baryon, carr2017primordial}.

\begin{equation}
    P_{\rm PBH}(m) = \frac{1}{\sqrt{2 \pi} \sigma_p m} \times \exp\bigg[-\frac{(\log(m/M_c))^2}{2 ~\sigma_{p}^{2}}\bigg],
\end{equation}
where $M_c$ is the characteristic mass scale and $\sigma_{p}$ is the standard deviation of the $\log(m/M_c)$. The log-normal distribution is motivated by the small-scale density fluctuations \citep{dolgov1993baryon, carr2017primordial,raidal2017gravitational}.

\subsection{SGWB from a population of stellar mass BHBs}

The SGWB is conventionally characterized by its energy density per unit logarithmic frequency interval, normalized by the critical energy density of the Universe. Mathematically,

\begin{equation}
\Omega_{\mathrm{GW}}(f) = \frac{1}{\rho_c c^{2}} \frac{d \rho_{\mathrm{GW}}}{d \ln f},
\end{equation}
where $f$ is the observer frame frequency, $\rho_{\mathrm{GW}}$ is the energy density of the GW background, and $\rho_c = 3 H_0^2/8 \pi G$ is the critical density of the Universe, $H_0$ denotes the Hubble constant, and $G$ is the universal gravitational constant.

In general, $\Omega_{\mathrm{GW}}(f)$ encapsulates contributions from all unresolved GW sources, and can include both cosmological and astrophysical origins. In the frequency band probed by ground-based detectors, the dominant contribution is expected to arise from a population of compact binary coalescences (CBCs).

The mean $\Omega_{\rm GW}(f)$ from a population of BHBs can be written as 
\citep{phinney2001practical,christensen2018stochastic}

\begin{equation}
    \begin{aligned}
     \overline{\Omega}_{\rm GW}(f) = \frac{1}{\rho_c c^2} &\int\limits^{m_{max}}_{m_{\rm min}} dm_1 \int\limits^{m_{max}}_{m_{\rm min}} dm_2 \int\limits_{{\rm z}_{\rm min}}^{\infty} f_r ~d{\rm z}~ \frac{dV_c}{d{\rm z}} \\
    &\times \bigg[\frac{\mathcal{R}_{\rm{GW}}(z,m_1,m_2)}{1+{\rm z}}\bigg]  \bigg[\frac{1+{\rm z}}{4 \pi d_{L}^{2} c} \frac{dE_{\rm{gw}}}{df_r} \bigg],
    \end{aligned}
    \label{SGWB}
\end{equation}
where $\mathcal{R}_{\rm{GW}}({\rm z},m_1,m_2)$ is the source frame merger rate of BHB per unit comoving volume 
($V_c$) with component masses $m_1$ and $m_2$ at redshift z, given by
\begin{equation}        \mathcal{R}_{\rm{GW}}({\rm z},m_1,m_2) = \mathcal{R}_{\rm x}({\rm z}) ~ P_{\rm x}(m_1) ~ P_{\rm x}(m_2),
\end{equation}
where $\mathcal{R}_{\rm x}$, and $P_{\rm x}$ represents the merger rate and the mass distribution with $\rm x \equiv ABH~ or~ PBH $, $d_L$ is the luminosity distance of the source, and $f_r$ = $f$(1+{\rm z}) is the source frame frequency, and ${\rm z}_{\rm min}$ denotes the maximum redshift up to which individual GW sources can be resolved by the detector. $\frac{dE_{\rm{gw}}}{df_r}$ is the energy emitted by the source per unit source frame frequency. Throughout this paper, we denote the merger rate simply as $\mathcal{R}({\rm z})$ when referring to it in general terms.

For the fiducial model considered in this analysis, the local merger rate is set to
$\mathcal{R}_{\rm ABH}(0) = \mathcal{R}_{\rm PBH}(0) = 20~\mathrm{Gpc^{-3}yr^{-1}}$. The model parameters are set to $\beta = 1.3$, $\gamma = 2.7$, $\kappa = 5.6$, $\rm z_p = 1.9$,
$\alpha = -2.3$, $\lambda = 0.1$, $\sigma_{m} = 5~M_{\odot}$, and $\sigma_{p} = 0.5~M_{\odot}$. The minimum redshift, $\rm z_{\min}$, is chosen to correspond to the median detection horizon over all source configurations, evaluated at the median mass of each population. The parameters $\gamma$ and $\kappa$ are chosen to reproduce a redshift evolution similar to the cosmic star formation rate described in \citep{madau2014cosmic}. The adopted local merger rate is consistent with the current observational constraints from the LVK Collaboration \citep{abac2025gwtc}. The value of parameter $\beta$ is consistent with most of the formation scenarios
of PBH with Poissonian clustering \citep{carr2017primordial,raidal2017gravitational, sasaki2018primordial,raidal2019formation}. We discuss the variation in our results by changing the model parameter in the range consistent with the results from GWTC-4 of the LVK Collaboration \citep{abac2025gwtc}.

\subsection{Spectral covariance of GW Background due to simulated population of the BHB}\label{Sec:Theory_Cov}

The number of BH mergers contributing to the SGWB within a given time interval is expected to follow a Poisson distribution \citep{dvorkin2018exploring,mukherjee2020time,sah2024non}. This naturally gives rise to fluctuations in the number of merger events between two different observation bins \citep{mukherjee2020time, Mukherjee:2020jxa, ginat2020probability, Braglia:2022icu, ginat2023frequency,sah2024non}. The amplitude of these fluctuations depends on both the intrinsic merger rate and the length of individual observation bins. A time-varying, or non-stationary, SGWB signal resulting from these fluctuations induces correlations between different frequency modes, causing the signal covariance matrix to acquire non-zero off-diagonal terms. The detailed structure of this spectral covariance matrix is expected to depend on the underlying mass distribution of the compact binary population contributing to the background, while the overall amplitude scales with the total merger rate \cite{sah2024non}. The spectral covariance between two frequency bins, $f_1$ and $f_2$, of the SGWB energy-density spectrum $\Omega_{\rm GW}(f)$ is defined as \citep{sah2024non}

\begin{equation}
\begin{aligned}
    \mathcal{C}_{S}(f_1, f_2) = &\Big\langle 
   \Big(\Omega_{\rm GW}(f_1) - \big\langle \Omega_{\rm GW}(f_1) \big\rangle\Big)\\
    & \times \Big( \Omega_{\rm GW}(f_2) - \big\langle \Omega_{\rm GW}(f_2) \big\rangle\Big)\Big\rangle,
\end{aligned}
\label{Covs}
\end{equation}
where $\langle \cdot \rangle$ denotes the ensemble average.

\begin{figure}
    \centering
   \subfigure[]{\label{CovMp40}
    \includegraphics[width=\linewidth]{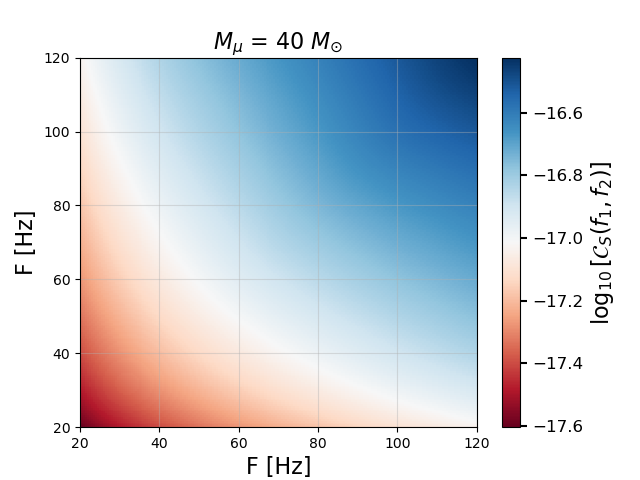}}
    \subfigure[]{\label{CovMp120}
    \includegraphics[width=\linewidth]{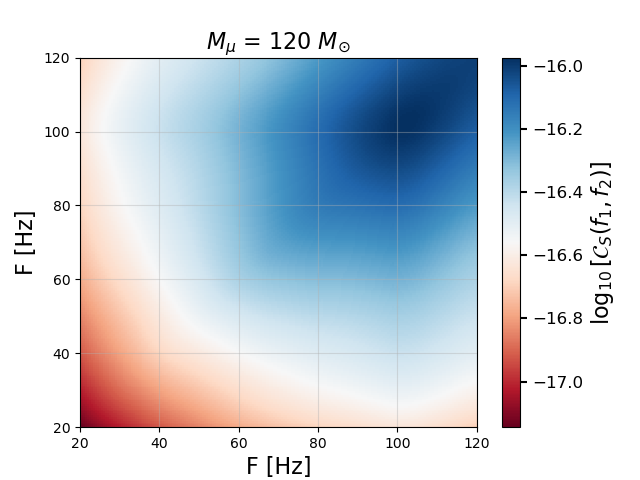}}
    \caption{Covariance matrix $\mathcal{C}_{S}(f_1, f_2)$ of $\Omega_{\rm GW}(f)$ (see Eq.~\ref{Covs}), computed over a time segment of $\Delta T = 192$ seconds. The covariance between different frequency modes is shown for the ABH population with (a) $M_{\mu}=40\,M_{\odot}$ and (b) $M_{\mu}=120\,M_{\odot}$. The color scale denotes $\log_{10}\big[\mathcal{C}_{S}(f_1, f_2)\big]$.}

\label{fig:Cs_ABH}   
\end{figure}

\begin{figure}
    \centering
    \subfigure[]{\label{CovMc40}
    \includegraphics[width=\linewidth]{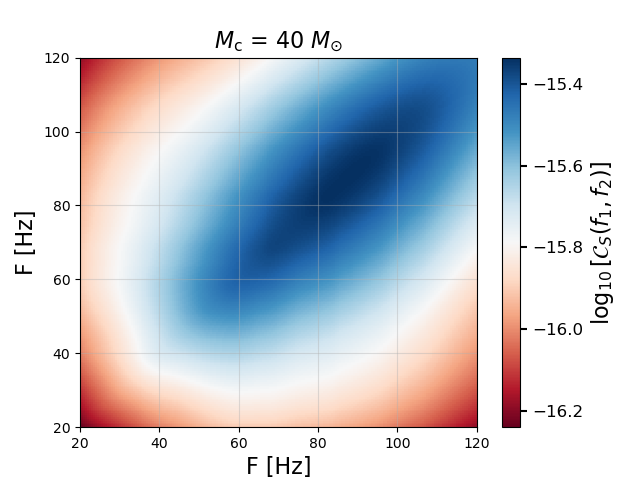}}
    
    \subfigure[]{\label{CovMc120}
    \centering
    \includegraphics[width=\linewidth]{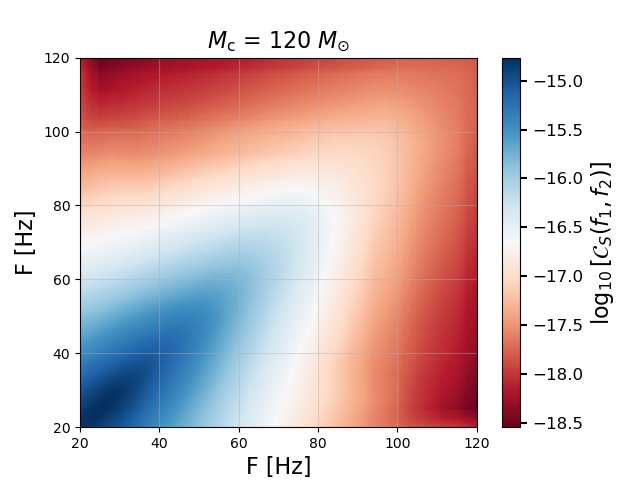}}
    \caption{Covariance matrix $\mathcal{C}_{S}(f_1, f_2)$ of $\Omega_{\rm GW}(f)$ (see Eq.~\ref{Covs}), computed over a time segment of $\Delta T = 192$ seconds. The covariance between different frequency modes is shown for the PBH population with (a) $M_{\rm c}=40\, M_{\odot}$ and (b) $M_{\rm c}=120\, M_{\odot}$. The color scale indicates $\log_{10}\big[\mathcal{C}_{S}(f_1, f_2)\big]$.}
\label{fig:Cs_PBH}    
\end{figure}

In Fig. \ref{fig:Cs_ABH} we show the spectral covariance structure of SGWB density due to a population of ABHs for two $M_{\mu}$: $M_{\mu} = 40 ~M_\odot$ and $M_{\mu} = 120~ M_\odot$. We set the minimum mass, $m_{\rm min}$, to $5 ~M_\odot$. All the other parameters are kept fixed to their fiducial values as discussed in the previous section. Similarly, Fig. \ref{fig:Cs_PBH} presents the covariance matrices for PBH populations with $M_{c} = 40~M_\odot$ and $M_{c} = 120~M_\odot$, with all other parameters fixed to their fiducial values. 

The structure of the covariance matrix is strongly influenced by the underlying mass distribution. In populations dominated by lower-mass BHs, GW emission spans a broader frequency range. As a result, the
structure of the covariance between frequency modes extends up to larger frequency separations. Conversely, for populations with higher-mass BHs, the covariance structure is stronger at lower frequencies than at higher frequencies. This behavior arises from the fact that the maximum frequency emitted during a binary merger is inversely proportional to the total mass of the system. Furthermore, the covariance structure differs significantly between ABH and PBH models. This distinction is primarily driven by differences in their mass distributions. PBH populations are modeled using log-normal (LN) distributions, whereas ABH populations are modeled by power-law + Gaussian (PLG) distributions. The PLG distribution peaks at the lower-mass end compared to the LN distribution, which peaks at higher mass and extends toward both sides. As a result, the covariance structure of PBHs exhibits sharper features.

These differences in the covariance structure encode additional information about the underlying population that is not captured by the power spectrum alone. The power spectrum is often insufficient to distinguish between different population properties due to inherent parameter degeneracies. In contrast, the spectral covariance matrix retains distinct imprints of features such as the mass distribution, thereby breaking these degeneracies. As a result, we focus on constraints derived from the spectral covariance in this analysis, as it provides significantly greater statistical power in probing population parameters \cite{sah2024non}.

\section{Methodology}\label{Meth}

In this section, we describe the procedure used to compute the cross-correlation of strain data from the Hanford (H1) and Livingston (L1)  detectors and to construct the spectral covariance matrix. Our analysis is based on publicly available strain data from the O3 and O4a observing runs.

\subsection{Cross-correlation between two detectors}

The standard approach to SGWB searches in ground-based detectors relies on cross-correlating strain data from two spatially separated interferometers \citep{romano2017detection,christensen2018stochastic,renzini2022stochastic}. The basic assumption is that a stochastic signal will be present in both detectors, while the instrumental noise, being uncorrelated between sites, averages out to zero over time. Let $s_I(t)$ and $s_J(t)$ denote the time series strain data from detectors $I$ and $J$, respectively. Each strain signal is modeled as a sum of the GW strain $h(t)$ and detector noise, $n(t)$,
\begin{equation}
s_I(t) = h_I(t) + n_I(t), \quad s_J(t) = h_J(t) + n_J(t).
\end{equation}
The cross-correlation estimator for the SGWB energy density spectrum $\Omega_{\mathrm{GW}}(f)$ is obtained by cross-correlating the strain between two detectors
\citep{abbott2023open,ligo2025open} 
\begin{equation}
\hat{C}_{IJ}(f; t) = \frac{1}{\Delta T} \frac{20 \pi^{2}}{3 H_{0}^{2} ~\gamma_{IJ}(f)} f^{3} ~\mathrm{\textbf{Re}}[\tilde{s}_I^*(f; t)\tilde{s}_J(f; t)],
\label{Cov_IJ}
\end{equation}
where $\tilde{s}_I(f; t)$ is the Fourier transform of the $s_I(t)$ over a segment of duration $\Delta T$ centered at time $t$. The operator $\mathrm{\textbf{Re}}$ denotes the real part, and $\gamma_{IJ}(f)$ is the overlap reduction function that encodes the relative orientation and separation between the two detectors \citep{thrane2009probing,abbott2021upper}.

\subsection{Spectral covariance estimator}
The estimator of the spectral covariance matrix, $\hat{\mathcal{C}}_{S,t}^{IJ}(f_1,f_2)$, of the $\Omega_{\rm GW}(f)$ for a time segment t can be written as 
\begin{equation}
    \begin{aligned}
        \hat{\mathcal{C}}_{S,t}^{IJ}(f_1,f_2)&=  \Big( \hat{C}_{IJ}(f_1; t) - \langle\hat{C}_{IJ}(f_1,t)\rangle\Big)  \\
        & \times \Big( \hat{C}_{IJ}(f_2; t) - \langle\hat{C}_{IJ}(f_2,t)\rangle\Big).
    \end{aligned}
\end{equation}
To obtain an optimal estimator of the spectral covariance of $\Omega_{\rm GW}(f)$, we average the individual segment estimators $\hat{\mathcal{C}}_{S,t}^{IJ}(f_1,f_2)$ across all available time segments, weighting each by the inverse of their corresponding variance $\hat{\Sigma}^{IJ}_{t}(f_1,f_2)$. The factor $\hat{\Sigma}^{IJ}_{t}(f_1,f_2)$ is given by

\begin{equation}
    (\hat{\Sigma}^{IJ}_{t}(f_1,f_2))^{2} =  \hat{\sigma}_{I}(f_1,t)  ~\hat{\sigma}_{J}(f_1,t)  ~\hat{\sigma}_{I}(f_2,t)  ~\hat{\sigma}_{J}(f_2,t),
\end{equation}
where
\begin{equation}
\hat{\sigma}_{I}(f; t) = \frac{1}{\Delta T} \frac{20 \pi^{2}}{3 H_{0}^{2} ~\gamma_{IJ}(f)} f^{3} ~\tilde{s}_I^*(f; t)\tilde{s}_I(f; t),
\label{Sigma_Cov_IJ}
\end{equation}
$\hat{\sigma}_{I}(f; t)$ represents the measured auto-correlation in the segment $t$. The mean $\langle(\hat{\Sigma}^{IJ}_{t}(f_1,f_2))^2\rangle$ is the variance in the measurement of $\langle\hat{\mathcal{C}}_{S,t}^{IJ}(f_1,f_2)\rangle$. The derivation of the noise, $\hat{\Sigma}^{IJ}_{t}(f_1,f_2)$,  is presented in Appendix. \ref{sec:Cov_nos}.

The noise-weighted estimator of the spectral covariance is given by
\begin{equation}
    \hat{\mathcal{C}}_{S,w}^{IJ}(f_1,f_2) = \frac{\sum\limits_{t} \hat{\mathcal{C}}_{S,t}^{IJ}(f_1,f_2)(\hat{\Sigma}^{IJ}_{t}(f_1,f_2))^{-2}}{\sum\limits_{t}(\hat{\Sigma}^{IJ}_{t}(f_1,f_2))^{-2}}.
    \label{Cov_weight_t}
\end{equation}
The standard deviation of $\hat{\mathcal{C}}_{S,w}^{IJ}(f_1,f_2)$ is given by

\begin{equation}
    (\hat{\Sigma}^{IJ}_{w}(f_1,f_2))^2 =\frac{1}{\sum\limits_{t}  \frac{1}{(\hat{\Sigma}^{IJ}_{t}(f_1,f_2))^2}}.
\end{equation}
The $\hat{\mathcal{C}}_{S,w}^{IJ}(f_1,f_2)$ serves as a weighted averaged estimator for the spectral correlation $\mathcal{C}_{S}(f_1, f_2)$.

\begin{figure*}
    \centering    \includegraphics[width=0.7\linewidth]{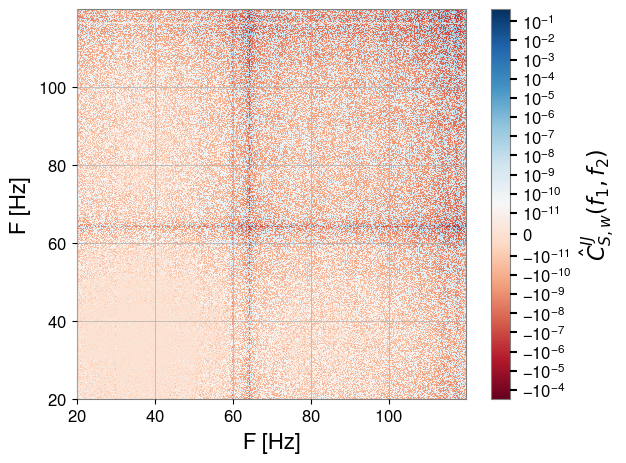}
    \centering
    \caption{The signal spectral covariance matrix $\hat{\mathcal{C}}_{S,w}^{IJ}(f_1, f_2)$ (see Eq. \ref{Cov_weight_t}) of the $\Omega_{\rm GW}(f)$ estimator obtained by cross-correlating the strain data from H1 and L1 detectors during O3 and O4a of the LVK observing runs.  The covariance is estimated as a weighted average of the spectral covariances over all short-time Fourier transform segments, with the data coarse-grained to a frequency resolution of $1/32~\mathrm{Hz}$ and a segment duration of $\Delta T = 192$ seconds. A symmetric logarithmic (symlog) color scale is used to display the wide dynamic range of $\hat{\mathcal{C}}_{S,w}^{IJ}(f_1, f_2)$ values, capturing both positive and negative correlations.}
\label{fig:Cs_data}    
\end{figure*}

\section{Results from LVK O3 and O4a observing runs}\label{Result}
For this analysis, we use publicly available strain data from the LIGO H1 and L1 detectors during the O3 and O4a \citep{abbott2023open,ligo2025open}. We use data that are downsampled to 2048 Hz and span the full duration of each run, excluding periods when one or both detectors were not operating in science mode.

To ensure data quality and reduce contamination from known instrumental artifacts, we apply a series of masks to exclude bad data segments. We first remove any segments that overlap with known GW events reported by the LVK collaboration. We then notch out the frequency bands that show significant lines that may contaminate the signal. We also apply non-stationarity cuts ($\Delta\sigma_{\rm cut}$) following the procedure adopted in previous LVK analyses \citep{abbott2023open, ligo2025open}. Specifically, we discard time segments for which the square root of the variance changes by more than $10\%$ between consecutive segments in at least one of the detectors. This threshold is more stringent than the $20\%$ criterion typically used in LVK analyses. We adopt this conservative choice because we find that a $20\%$ threshold retains a few contaminated segments (with duration $\Delta T = 192$ seconds) that exhibit spurious correlations between frequencies in the estimated $\hat{C}_{IJ}(f; t)$. These segments introduce excess power and spectral artifacts that are inconsistent with the expectation from compact object mergers.  We interpret these features as likely arising from some residual correlated noise (e.g., Schumann resonances \cite{schumannUberDampfungElektromagnetischen1952}) between detectors rather than a genuine astrophysical signal \citep{PhysRevD.107.022004, thrane2013correlated,thrane2014correlated,coughlin2018measurement,janssens2021impact}. By imposing a stricter $10\%$ cut, we effectively suppress such contamination, ensuring that the remaining data better yield a more reliable estimate of the cross-correlation statistic. This choice does not significantly reduce the sensitivity, while improving the robustness of the inferred spectrum.

After applying these cuts, the remaining data are divided into time segments of duration $\Delta T = 192$ seconds. Each segment is windowed using a Planck-taper window before performing a discrete Fourier transform (DFT), yielding the frequency-domain strain $\tilde{s}_I(f; t)$ for each detector $I = \{ \mathrm{H1}, \mathrm{L1} \}$. We then compute the cross-correlation estimator $\hat{C}_{IJ}(f; t)$ in each segment, and construct the spectral covariance matrix from these values as described in the previous subsection.

\subsection{Spectral Covariance from LIGO Data}

Fig. \ref{fig:Cs_data} presents the noise-weighted estimator of the spectral covariance matrix, $\hat{\mathcal{C}}_{S,w}^{IJ}(f_i, f_j)$, computed using strain data from the O3 and O4a observing runs. The correlation is obtained between the frequency bins coarse-grained at a frequency resolution of 1/32 Hz. Given the wide dynamic range of $\hat{\mathcal{C}}_{S,w}^{IJ}(f_1, f_2)$—with fluctuations spanning both positive and negative values—we adopt a symmetric logarithmic (symlog) color scale to visualize the full structure effectively. 

We note that, unlike the theoretical spectral covariance matrices shown in Figs.~\ref{fig:Cs_ABH} and \ref{fig:Cs_PBH}, which are strictly positive, the observational estimator $\hat{\mathcal{C}}^{IJ}_{S,w}(f_1,f_2)$ in Fig.~\ref{fig:Cs_data} exhibits both positive and negative values. In the theoretical modeling, the spectral covariance is defined as an ensemble average over realizations of the SGWB signal and captures correlations induced by a population of merging binaries. In contrast, the observational covariance is a noise-dominated, finite-duration estimator obtained by cross-correlating strain data from two detectors (see Eq.~\eqref{Cov_IJ}), which naturally allows both positive and negative correlations.

As expected, the amplitude of fluctuations increases above 60 Hz, reflecting the increase in noise level due to the frequency-dependent suppression by the overlap reduction function (ORF) between the H1 and L1 detectors. 

In addition to stationary noise, a non-stationary noise budget contributes to residual correlations in the data \citep{vajente2020machine,zackay2021detecting,mozzon2022does}. Temporal variations 
due to instrumental and environmental conditions can introduce fluctuations in the strain noise level that vary across time segments. These effects can produce weak off-diagonal features in the noise covariance matrix. To mitigate these contributions, we apply a non-stationarity cut as mentioned in the last subsection.

\subsection{Chi-squared value of the data}

\begin{figure*}
    \centering    \includegraphics[width=0.7\linewidth]{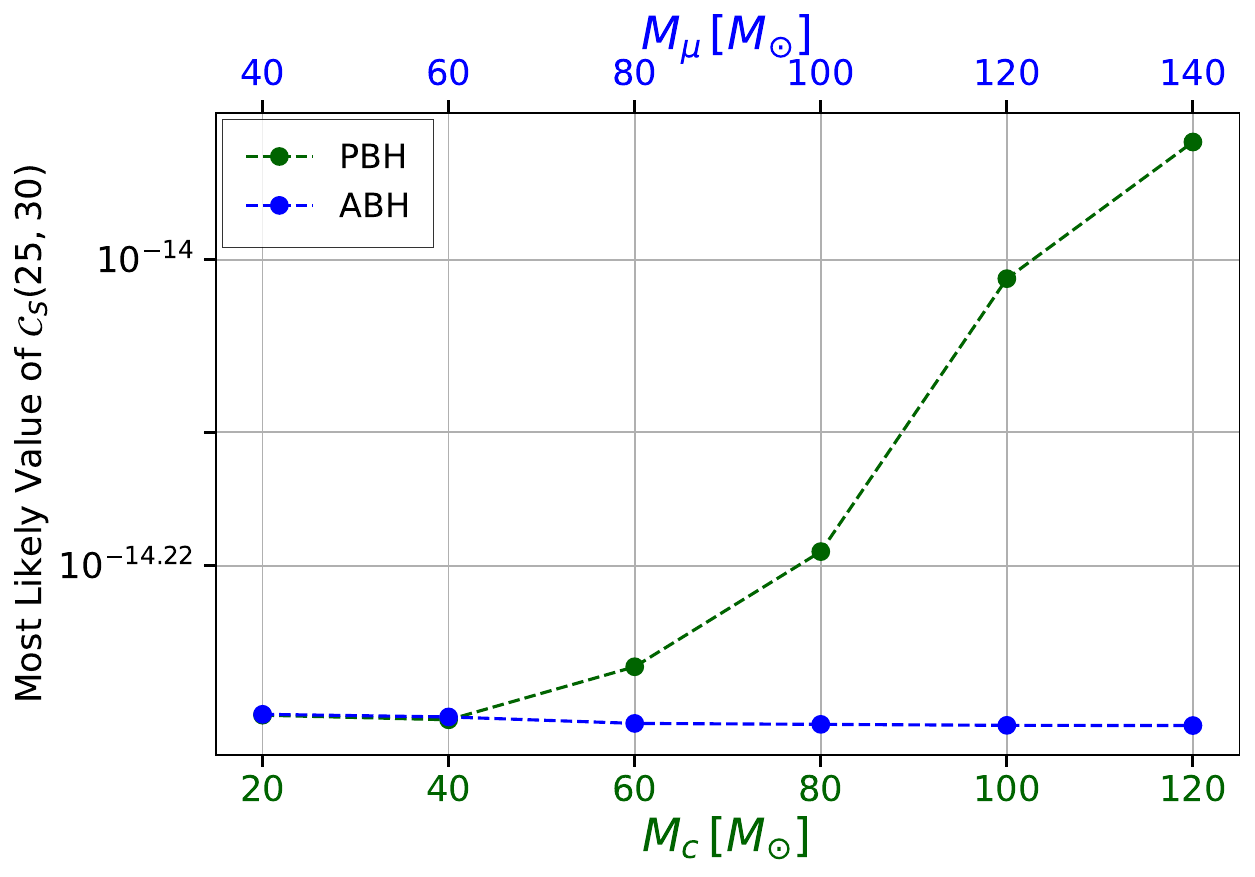}
    \caption{Most likely value of the spectral covariance amplitude, 
    $\mathcal{A}_{\rm min}\,\mathcal{C}^{\rm model}_{S}(25,30)$, 
    corresponding to the value of $\mathcal{A}$ that minimizes 
    $\chi^{2}_{\rm model}$, shown as a function of $M_{\mu}$ for the ABH population 
    and $M_{c}$ for the PBH population, with all other parameters fixed to their fiducial values. 
    This figure illustrates how the best-fit spectral covariance amplitude varies 
    with the characteristic mass scale of the binary population. 
    For the ABH models, the inferred amplitude remains approximately constant over the range of $M_{\mu}$ considered, whereas for PBH models, 
    it exhibits a systematic increase with $M_{c}$.}
    \label{fig:Max_Like}  
\end{figure*}

We do not perform Bayesian inference on astrophysical model parameters in this work, as the current data are not expected to contain a detectable signal that would enable meaningful parameter constraints. 
Instead, we focus on searching for signatures of a stochastic signal. We compare the features observed in the data with expectations derived from simulated BHB populations. To quantify the agreement between the empirical covariance matrix and the model predictions, we define a chi-square-like statistic as

\begin{equation}
\chi^2_{\rm model} = \sum_{f_1,f_2< f_1} \frac{\left| \hat{\mathcal{C}}_{S,w}^{IJ}(f_1, f_2) - \mathcal{A}~ 
\mathcal{C}_S^{\rm model}(f_1, f_2,\theta)\right|^2}{\big(\hat{\Sigma}^{IJ}_{w}(f_1,f_2)\big)^2},
\end{equation}
where $\hat{\mathcal{C}}_{S,w}^{IJ}(f_1, f_2)$ is the covariance matrix defined in Eq. \eqref{Cov_weight_t}, and $\mathcal{C}_S^{\rm model}(f_1, f_2,\theta)$ is the predicted covariance from a given population model with population parameter $\theta$ and it acts as the template of the covariance matrix.
$\mathcal{A}$ sets the overall scaling of the covariance matrix, and its value depends on the choice of the model template. We exclude the diagonal terms ($f_1 = f_2$) from the summation, as they correspond to the auto-power components, for which $\hat{\mathcal{C}}_{S,w}^{IJ}(f_1, f_1)$ acts as a biased estimator of the true covariance structure due to noise.

We fix all parameters to their fiducial values and vary only $M_{\mu}$ in the case of ABHs and $M_{c}$ for PBHs. In Fig. \ref{fig:Max_Like}, we plot the most likely value of the spectral covariance between frequencies 25 Hz and 30 Hz, $\mathcal{A_{\rm min}}\times\mathcal{C}_{S}^{\rm model}(25,30)$, corresponding to the values of $\mathcal{A}$ that minimizes $\chi^{2}_{\rm model}$ ($\mathcal{A}_{\rm min}$)  as a function of $M_\mu$ (ABH) and $M_{c}$ (PBH). The most likely value shows modest dependence on the mass distribution parameter for the ABH population. The most likely values of PBHs, in general, show higher values when compared to the ABH models.

\subsection{Template-based spectral covariance search technique}\label{template}

\begin{figure*}
    \centering    \includegraphics[width=0.7\linewidth]{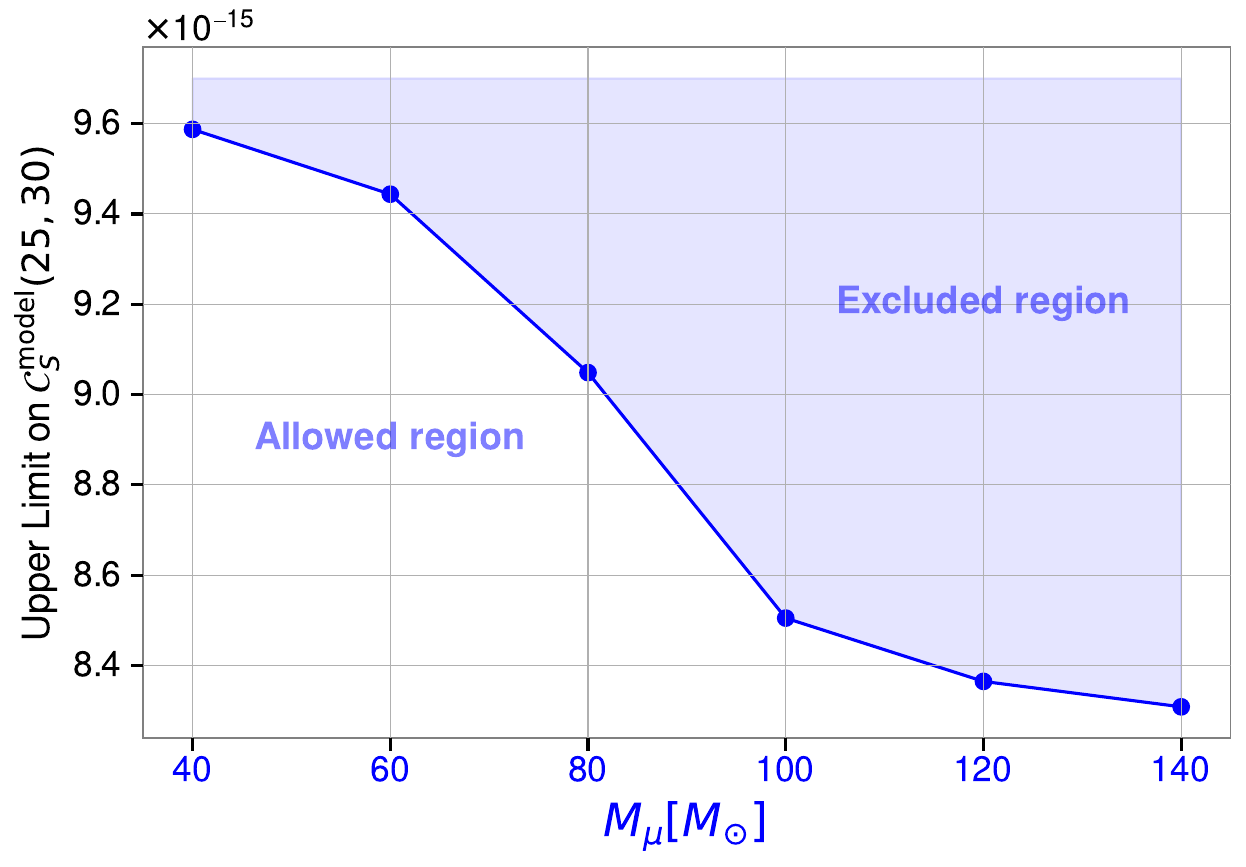}
    \caption{Template-dependent upper limits on the spectral correlation in the SGWB energy density spectrum, 
    $\Omega_{\rm GW}(f)$, for the ABH population, represented here as the covariance between 25~Hz and 30~Hz, as a function of $M_{\mu}$, with all other parameters fixed to their fiducial values. 
    Each point corresponds to the 95\% credible upper limit on the spectral covariance amplitude, 
    with the corresponding optimal estimates reported in Table~\ref{tab:point_estimates}. 
    The upper limits are derived using a model-based template-weighted estimator that accounts for the expected spectral correlation structure of the signal.}
    \label{fig:Upper_ABH}  
\end{figure*}

\begin{figure*}
    \centering    \includegraphics[width=0.7\linewidth]{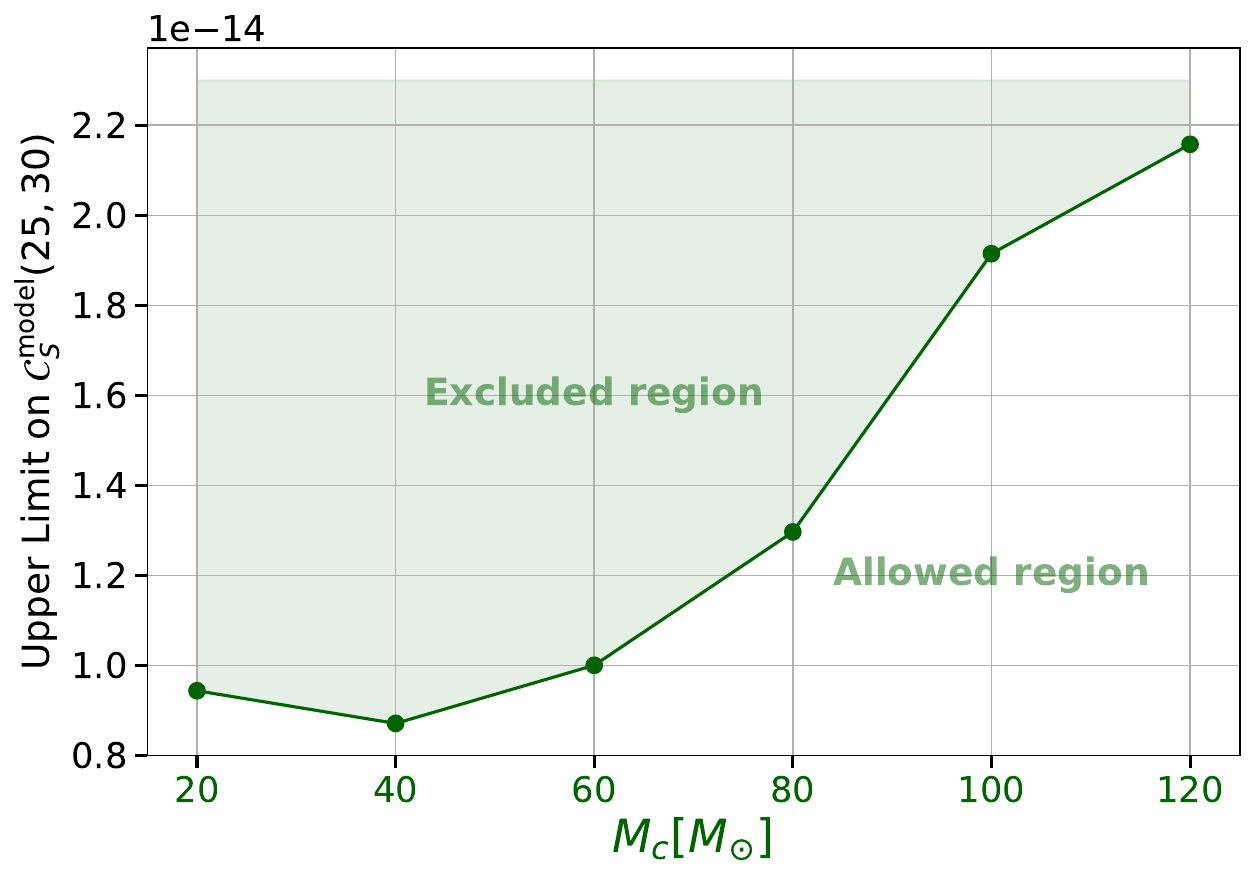}
    \caption{Template-dependent upper limits on the spectral correlation in the SGWB energy density spectrum, 
    $\Omega_{\rm GW}(f)$, for the PBH population, represented here as the covariance between 25~Hz and 30~Hz, as a function of $M_{c}$, with all other parameters fixed to their fiducial values. 
    Each point corresponds to the 95\% credible upper limit on the spectral covariance amplitude, 
    with the corresponding optimal estimates reported in Table~\ref{tab:point_estimates}. 
    The upper limits are derived using a model-based template-weighted estimator that accounts for the expected spectral correlation structure of the signal.}
    \label{fig:Upper_PBH}  
\end{figure*}

We use the simulated templates of the spectral covariance from the astrophysical models (as obtained in Sec. \ref{Sec:Theory_Cov}) to define an optimal estimator by taking a weighted average of the measured spectral covariance ($\hat{\mathcal{C}}_{S,w}^{IJ}(f_1, f_2)$) over all frequency pairs, weighted by the simulated spectral covariance, and the corresponding standard deviation. The template-based optimal estimator of the spectral covariance is given by

\begin{equation}
    \hat{\mathcal{C}}_{S,\rm opt}^{IJ} \equiv \mathcal{C}^{\rm model}_{S}{\scriptstyle(f^{\rm ref}_{1}, f^{\rm ref}_{2},\theta)}\,\frac{\sum\limits_{f_1, f_2 < f_1} \hat{\mathcal{C}}_{S,w}^{IJ}(f_1,f_2) \Big[\frac{\mathcal{C}^{\rm model}_{S}(f_1, f_2,\theta)}{(\hat{\Sigma}^{IJ}_{w}(f_1,f_2))^{2}}\Big]}{\sum\limits_{f_1, f_2 < f_1}\Big(\frac{\mathcal{C}^{\rm model}_{S}(f_1, f_2,\theta)}{\hat{\Sigma}^{IJ}_{w}(f_1,f_2)}\Big)^{2}},
\end{equation}

\begin{equation}
    \hat{\Sigma}^{IJ}_{\rm opt}(f_1,f_2) =\frac{\mathcal{C}^{\rm model}_{S}(f^{\rm ref}_{1}, f^{\rm ref}_{2})}{\sqrt{\sum\limits_{f_1,f_2 < f_1}\Big(\frac{\mathcal{C}^{\rm model}_{S}(f_1, f_2)}{\hat{\Sigma}^{IJ}_{w}(f_1,f_2)}\Big)^{2}}},
\end{equation}
where $\hat{\Sigma}^{IJ}_{\rm opt}(f_1,f_2)$ is the standard deviation of the optimal estimator of the spectral
covariance ($\hat{\mathcal{C}}_{S,\rm opt}^{IJ} $). The normalization factor $\mathcal{C}^{\mathrm{model}}_{S}(f^{\rm ref}_{1}, f^{\rm ref}_{2})$ is chosen such that the estimator $\hat{\mathcal{C}}^{IJ}_{S,\,\mathrm{opt}}$ represents the spectral covariance between the two reference frequencies $(f^{\rm ref}_{1}, f^{\rm ref}_{2})$. 
This normalization ensures that the optimal estimator preserves the same dimensional scaling as the empirical covariance, allowing for a direct comparison between model predictions and the data.

\begin{figure*}
    \centering    \includegraphics[width=0.7\linewidth]{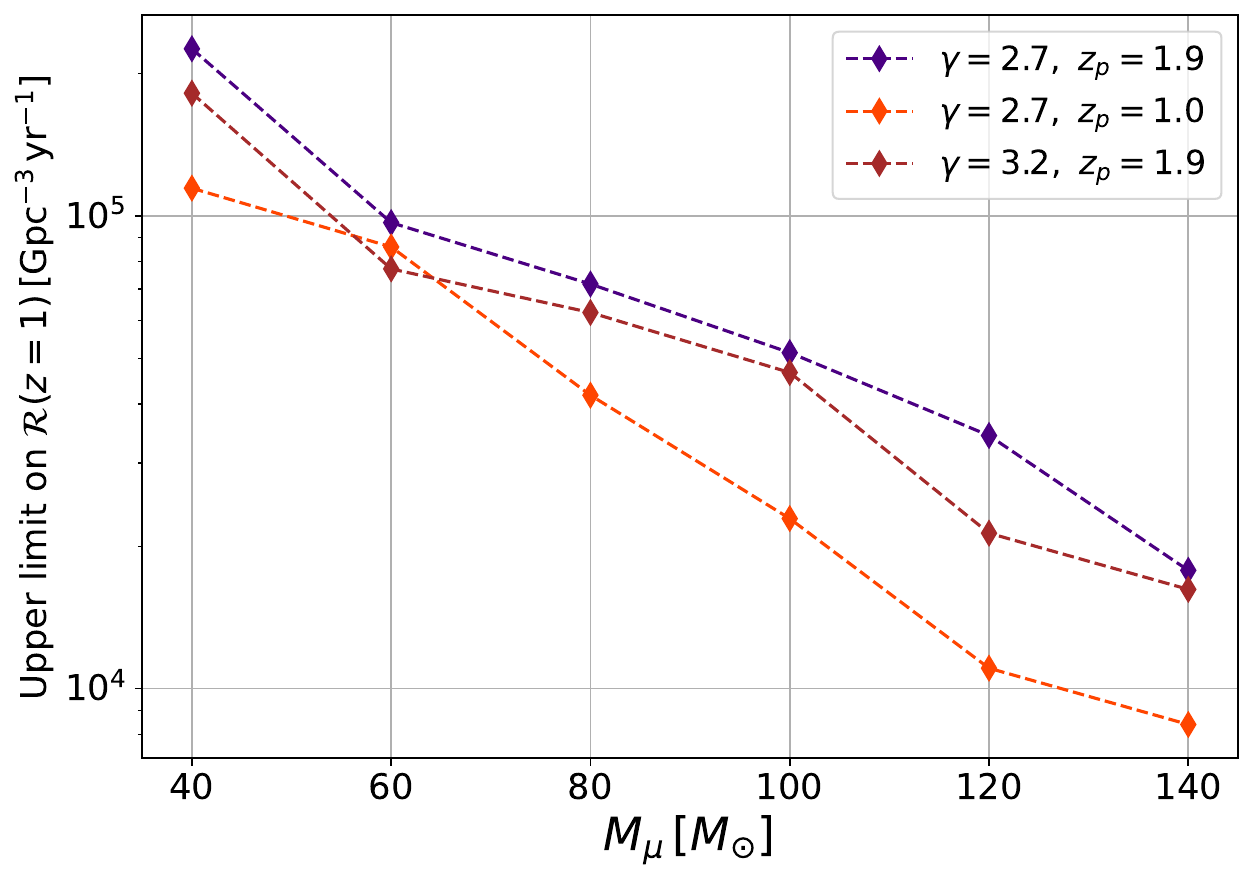}
    \caption{Upper limits on the merger rate of ABH at $z = 1$ as a function of the characteristic mass scale, $M_{\mu}$, for different values of $\gamma$ and $\rm z_p$, keeping the remaining parameters fixed to their fiducial values. These constraints are derived from the upper bound on the spectral covariance obtained using the template-based optimal estimator.}
    \label{fig:R_M_ABH}  
\end{figure*}

\begin{figure*}
    \centering    \includegraphics[width=0.7\linewidth]{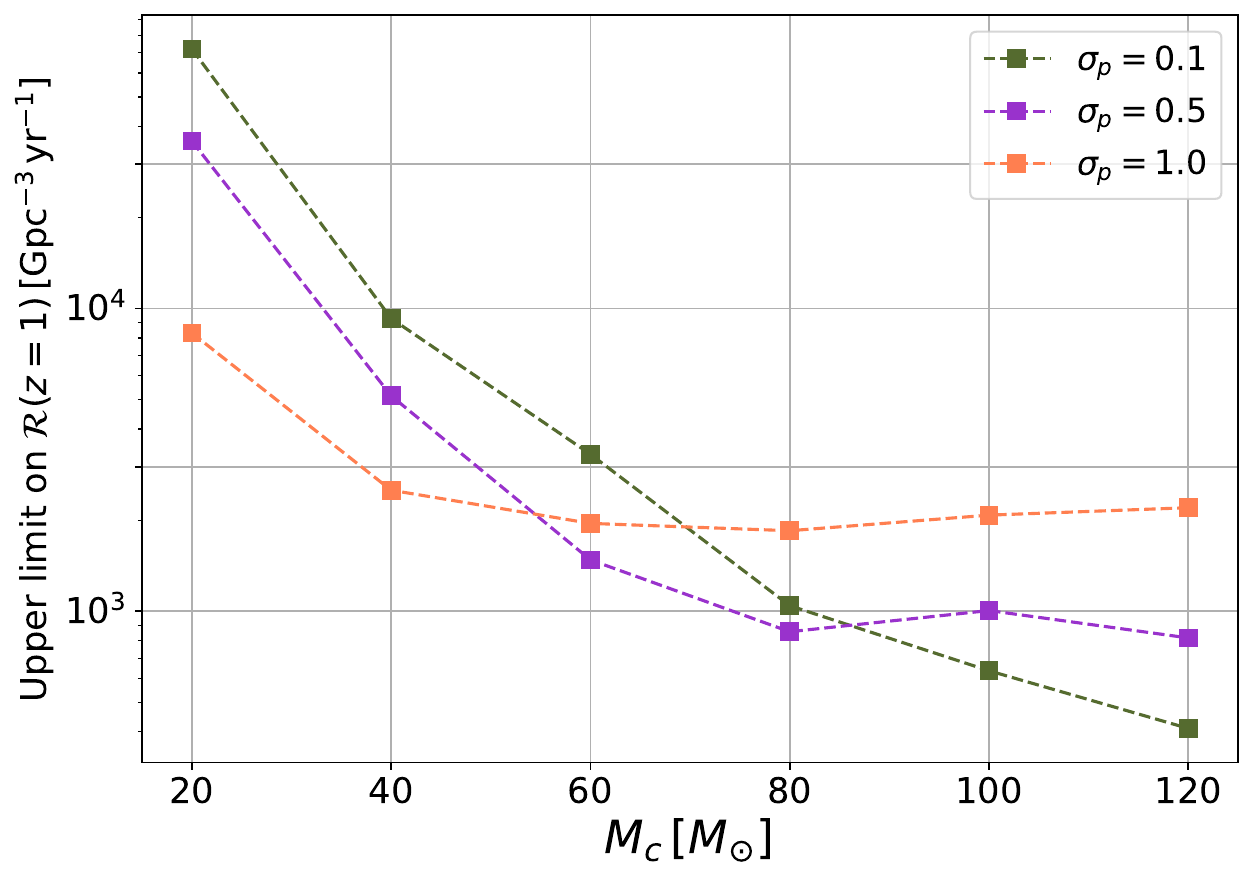}
    \caption{Upper limits on the merger rate of PBH at $z = 1$ as a function of the characteristic mass scale, $M_{c}$, for for three different values of $\sigma_p$, keeping the remaining parameters fixed to their fiducial values. These constraints are derived from the upper bound on the spectral covariance obtained using the template-based optimal estimator.}
    \label{fig:R_M_PBH}  
\end{figure*}

We employ this optimal estimator to search for signatures of spectral covariance in the data using a set of model-based templates. For computational efficiency, we restrict the analysis to frequencies $\leq 100~\mathrm{Hz}$. This cutoff has a negligible impact, as it retains more than $99\%$ of the total signal sensitivity for typical astrophysical sources \citep{abbott2023open, ligo2025open}. We express the optimal estimator in terms of the spectral correlation between 25~Hz and 30~Hz (i.e., $f^{\rm ref}_{1} = 25~\mathrm{Hz}$ and $f^{\rm ref}_{2} = 30~\mathrm{Hz}$).

We do not find any statistically significant evidence for a covariance structure in the data. 
In the absence of such evidence, we report the upper limits on the amplitude of the covariance structure. In the table. \ref{tab:point_estimates}, we report the optimal estimates of $\hat{\mathcal{C}}_{S,\rm opt}^{IJ}$ along with their $1 \sigma$ uncertainties.

In Fig. \ref{fig:Upper_ABH} and Fig. \ref{fig:Upper_PBH}, we show the \textit{95\% credible upper limit} on the amplitude of the spectral covariance for a set of model-based templates of ABH and PBH populations, respectively. 
For ABHs, the upper limits range between $8.3\times10^{-15}$ and $9.5\times10^{-15}$ for $M_{\mu}$ between $40~M_{\odot}$ and $140~M_{\odot}$. 
Similarly, for PBHs, the limits vary between $8.7\times10^{-15}$ and $21\times10^{-15}$ for $M_{c}$ in the range $20~M_{\odot}$ and $120~M_{\odot}$.

We translate the upper limit on the amplitude of the spectral covariance into an upper bound on the merger rate of BHBs. In Fig.~\ref{fig:R_M_ABH}, we present the resulting upper limits on the high-redshift merger rate of ABH, expressed as merger rate at $\rm z = 1$, as a function of the characteristic mass scale $M_{\mu}$ for different values of $\gamma$ and $\rm z_p$, which are within the current measurement uncertainties from LVK data \cite{abac2025gwtc, abac2025upper}, with all other parameters fixed to their fiducial values. Similarly, in Fig.~\ref{fig:R_M_PBH}, we show the high-redshift merger rate of PBH as a function of the characteristic mass scale $M_{c}$ for three different values of $\sigma_p$. The upper limit on $\mathcal{R}(z = 1)$ is obtained by assuming the appropriate scaling of the spectral covariance matrix amplitude with the merger rate.

We find that, for both ABHs and PBHs, the upper limit on $\mathcal{R}(z = 1)$ decreases with increasing mass. For PBHs, however, this trend gradually flattens at higher mass scales. This behavior can be understood as follows. More massive binaries produce a larger spectral covariance signal, which requires a lower merger rate to remain consistent with the observed upper limits. However, as the mass increases further, the GW signal progressively shifts out of the detector sensitivity band, reducing the observed covariance amplitude and leading to a flattening of the constraints. For PBHs, the dependence on the width parameter $\sigma_p$ introduces additional features. At low values of $M_c$, larger $\sigma_p$ leads to smaller values of $\mathcal{R}(z = 1)$, since a broader mass distribution includes a greater contribution from heavier binaries that enhance the signal. In contrast, at higher $M_c$, this trend reverses because the contribution from very massive binaries increasingly shifts out of the detector band, suppressing the observable signal. Quantitatively, the upper limit on the ABH merger rate at $\rm z = 1$ ranges from $2.25 \times 10^{5}$ to $8.4 \times 10^{3}\,\mathrm{Gpc}^{-3}\,\mathrm{yr}^{-1}$ for $M_{\mu}$ between $40\,M_{\odot}$ and $140\,M_{\odot}$. For the PBH case, the corresponding limits vary from $7.2 \times 10^{4}$ to $4.1 \times 10^{2}\,\mathrm{Gpc}^{-3}\,\mathrm{yr}^{-1}$ for $M_c$ between $20\,M_{\odot}$ and $120\,M_{\odot}$.

\begin{figure*}
    \centering    \includegraphics[width=0.7\linewidth]{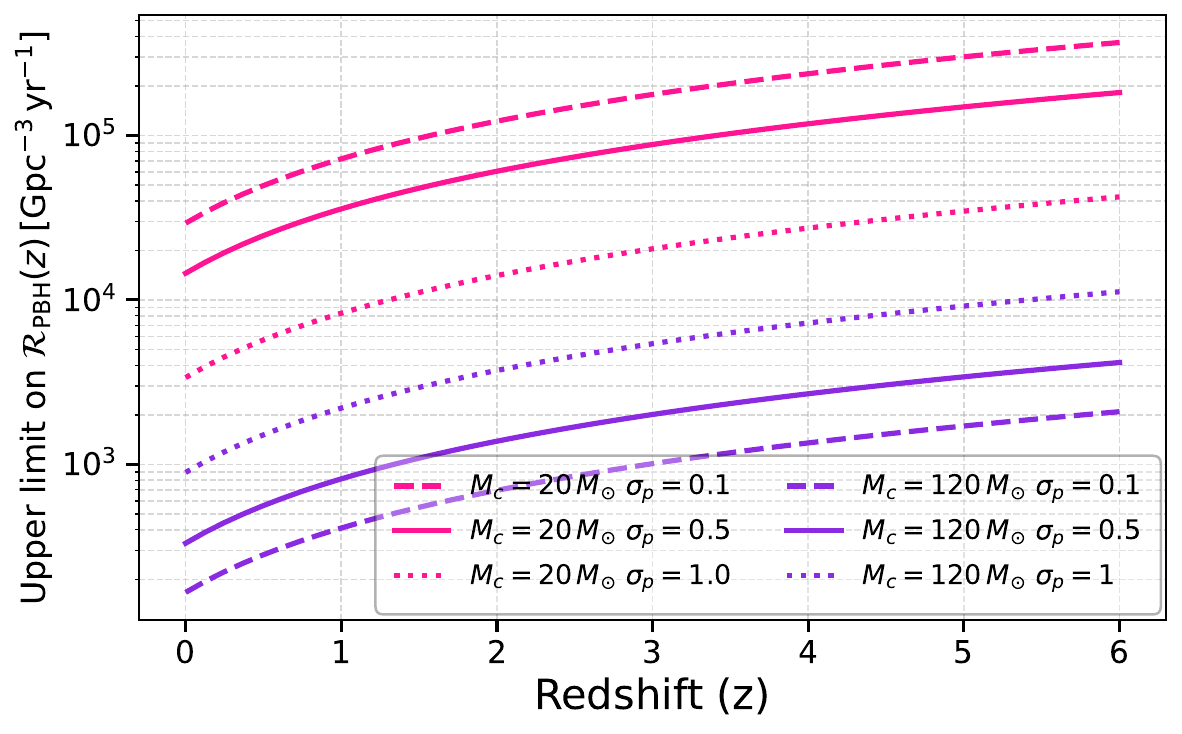}
    \caption{Upper limit in the merger rate as a function of redshift for PBH population for different values of $M_{c}$ and $\sigma_p$, with all other parameters fixed to their fiducial values. The upper limit increases with the redshift as the merger rate, given by Eq. \eqref{eq:R_PBH}, of the PBH increases with the redshift.}
    \label{fig:R_Z}  
\end{figure*}

 In Fig. \ref{fig:R_Z}, we show the corresponding upper limits on the merger rate as a function of redshift for different values of $M_{c}$ and $\sigma_p$ of PBH populations. The merger rate increases with redshift, leading to a corresponding rise in their upper limits. We avoid translating this bound to a dark matter fraction, as it becomes a model-dependent statement due to our ignorance of the suppression factor ($f_{\rm sup}$) and spatial clustering ($\xi_{\rm PBH}$).

The null detection of frequency–frequency correlations indicates that the SGWB observed in the O3 and O4a data is consistent with a stationary Gaussian process, with no evidence for intrinsic temporal fluctuations or non-stationary features. These findings establish a new observational benchmark for future searches. This analysis extends previous constraints on PBHs derived from O3 data using the SGWB power spectrum \cite{mukherjee2021can} by instead exploiting the non-stationary nature of the signal. Future observing runs with enhanced sensitivity and longer observation time will be essential to detect subtle spectral correlations and distinguish between competing population models \cite{Mukherjee:2021itf}.

\begin{table}
\centering
\caption{Optimal estimates of $\hat{\mathcal{C}}_{S,\rm opt}^{IJ}$ for different templates of ABH and PBH populations along with their measurement uncertainties.}
\begin{tabular}{|c|c||c|c|}

\hline
$M_{\mu}$/$M_{\odot}$ (ABH) & $\hat{\mathcal{C}}_{S,\rm opt}^{IJ}$/$10^{-15}$ & $M_{c}$$M_{\odot}$ (PBH) & $\hat{\mathcal{C}}_{S,\rm opt}^{IJ}/10^{-15}$ \\
\hline
40  &  4.07 $\pm$ 4.79  &  20  & 4.04 $ \pm$ 4.71 \\
60   &  4.0 $\pm$ 4.72  &  40  & 3.97 $ \pm$ 4.35  \\
80  &  3.93 $\pm$ 4.52  &  60 & 4.68 $\pm$ 5.01 \\
100 &  3.95 $\pm$  4.25  &  80 & 6.23 $\pm$ 6.48  \\
120  &  3.97 $\pm$ 4.18  &  100 & 9.77 $\pm$ 9.57 \\
140 &  3.99 $\pm$ 4.15  &  120 & 11.4 $\pm$ 10.7 \\
\hline
\end{tabular}
\label{tab:point_estimates}
\end{table}

\section{Conclusion}\label{conc}

In this study, we developed and implemented a framework to extract and characterize the spectral covariance structure of the SGWB from LVK strain data. By cross-correlating the strain from the Hanford (H1) and Livingston (L1) detectors from the O3 and O4a observing runs and evaluating the spectral covariance of this estimator across frequency bins, we go beyond the conventional power spectrum analysis to reveal additional statistical features of the SGWB.

Our results show that the measured spectral covariance matrix is consistent with noise, with no statistically significant spectral correlations. This is in line with expectations for a background dominated by uncorrelated Gaussian noise, with no clear evidence of frequency-frequency correlations.
Therefore, using a model-based template-weighted estimator, we place the first upper limits on the amplitude of the spectral covariance (between frequencies 25 Hz and 30 Hz) of the dimensionless SGWB energy density, $\Omega_{\rm GW}(f)$. The inferred upper limits are in the range $(8.3~-~21.5)\times10^{-15}$ for the
model templates considered in this work. Furthermore, we derive mass-scale–dependent upper limits on the merger rate. The upper limit on the high-redshift merger rate (at ${\rm z} = 1$) for the PBH population ranges from $7.2 \times 10^{4}$ to $4.1 \times 10^{2}\,\mathrm{Gpc}^{-3}\,\mathrm{yr}^{-1}$ as the median mass of the log-normal distribution varies between $20~M_{\odot}$ and $120~M_{\odot}$, for the range of mass distribution width parameters $\sigma_p$ considered in this analysis. These constraints are derived using only SGWB data, which carries the information from the high redshift Universe and hence brings complementary constraints than from GWTC-4 \cite{abac2025gwtc}.

This study provides a foundation for future investigations of the spectral properties of SGWB beyond conventional power-spectrum analyses. With improved strain sensitivity in upcoming detectors such as Cosmic Explorer \citep{hall2021gravitational,evans2021horizon} and the Einstein Telescope \citep{punturo2010einstein,maggiore2020science}, this method is expected to offer enhanced sensitivity to subtle spectral correlation features. Detecting or constraining such correlations would provide valuable insights into the statistical nature and the source of the SGWB.

\section*{Acknowledgments}
The authors are grateful to Shivaraj Kandhasamy for carefully reviewing the manuscript and providing useful comments as a part of the LIGO Publication and Presentation Policy. This work is a part of the $\langle \texttt{data|theory}\rangle$ \texttt{Universe-Lab}, which is supported by the TIFR and the Department of Atomic Energy, Government of India. This research is supported by the Prime Minister Early Career Research Award, Anusandhan National Research Foundation, Government of India. The authors would like to thank the  LIGO/Virgo scientific collaboration for providing the GW strain data. LIGO is funded by the U.S. National Science Foundation. Virgo is funded by the French Centre National de Recherche Scientifique (CNRS), the Italian Istituto Nazionale della Fisica Nucleare (INFN), and the Dutch Nikhef, with contributions by Polish and Hungarian institutes. This material is based upon work supported by NSF’s LIGO Laboratory, which is a major facility fully funded by the National Science Foundation. The authors would also like to acknowledge the use of the following Python packages in this work: Numpy \citep{van2011numpy,2020NumPy-Array}, Scipy \citep{jones2001scipy,virtanen2020scipy}, Matplotlib \citep{hunter2007matplotlib}, Astropy \citep{robitaille2013astropy,price2018astropy}, Ray \citep{moritz2018ray}, GWpy \citep{macleod2021gwpy}, and Pandas \citep{mckinney2011pandas}.
 
 \section*{Data Availability}
The data used in this work are publicly available through the Gravitational Wave Open Science Center (GWOSC) \citep{KAGRA:2023pio,Abac2025O4aOpenData}.

\bibliographystyle{apsrev4-2}
\bibliography{main.bib}

\appendix
\section{Derivation of spectral covariance noise} \label{sec:Cov_nos}

The Stochastic gravitational wave background (SGWB) is defined as gravitational wave (GW) energy density per unit logarithmic frequency interval, normalized by the critical energy density of the Universe. Mathematically, 
\begin{equation}
\Omega_{\mathrm{GW}}(f) = \frac{1}{\rho_c c^{2}} \frac{d \rho_{\mathrm{GW}}}{d \ln f},
\end{equation}
where $\rho_{\mathrm{GW}}$ is the energy density of the GW background, and $\rho_c  = \frac{3 H_0^2}{8 \pi G} $ is the critical density of the Universe.

We can define the spectral covariance between two frequencies in the $\Omega_{\rm GW}(f)$ as 
\begin{equation}
\begin{aligned}
    \mathcal{C}_{S}(f_1, f_2) = &\Big\langle 
   \Big(\Omega_{\rm GW}(f_1) - \big\langle \Omega_{\rm GW}(f_1) \big\rangle\Big)\\
    & \times \Big( \Omega_{\rm GW}(f_2) - \big\langle \Omega_{\rm GW}(f_2) \big\rangle\Big)\Big\rangle. 
\end{aligned}
\end{equation}

The estimator for the $\Omega_{\rm GW}(f)$ is obtained by cross-correlating the strain data from two widely separated detectors and is given by
\begin{equation}
\hat{C}_{IJ}(f; t) = \frac{1}{\Delta T} \frac{20 \pi^{2}}{3 H_{0}^{2} ~\gamma_{IJ}(f)} f^{3} ~\mathrm{\textbf{Re}}[\tilde{s}_I^*(f; t)\tilde{s}_J(f; t)]
\end{equation}
where $\tilde{s}_I^*(f; t)$ is the fourier transform of the strain $s_I(t) = h_I(t) + n_I(t)$ with $h_I(t)$ and $n_I(t)$ being the time-domain GW strain and noise, respectively. The Fourier transform is performed over a segment of duration $\Delta T$ centered at time $t$. The operator $\mathrm{\textbf{Re}}$ denotes the real part, $\gamma_{IJ}(f)$ is the overlap reduction function, and $H_0$ represents the Hubble constant.

The spectral covariance, $\mathcal{C}_{S}(f_1, f_2)$, can be written as
\begin{equation}
\begin{aligned}
     \mathcal{C}_{S}(f_1, f_2)  &= ~\Big\langle\hat{\mathcal{C}}_{S,t}^{IJ}(f_1,f_2)\Big\rangle\\
     &= \Bigg\langle 
    \Big( \hat{C}_{IJ}(f_1; t) - \Big\langle \hat{C}_{IJ}(f_1; t) \Big\rangle \Big)\\
    & \times \Big( \hat{C}_{IJ}(f_2; t) - \Big\langle \hat{C}_{IJ}(f_2; t) \Big\rangle \Big)
    \Bigg\rangle \\[4pt]
    &\equiv 
    \Bigg\langle 
    \Big( \tilde{s}_I^{*}(f_1, t)\tilde{s}_J(f_1, t) - 
    \Big\langle \tilde{s}_I^{*}(f_1, t)\tilde{s}_J(f_1, t) \Big\rangle \Big) \\
    &\times
    \Big( \tilde{s}_I^{*}(f_2, t)\tilde{s}_J(f_2, t) - 
    \Big\langle \tilde{s}_I^{*}(f_2, t)\tilde{s}_J(f_2, t) \Big\rangle \Big)
    \Bigg\rangle \\[4pt]
    &\approx 
    \Bigg\langle 
    \Big( \tilde{h}_I(f_1, t)\tilde{h}_J(f_1, t) - \Big\langle \tilde{h}_I(f_1, t)\tilde{h}_J(f_1, t) \Big\rangle \Big)\\
    &\times\Big( \tilde{h}_I(f_2, t)\tilde{h}_J(f_2, t) - \Big\langle \tilde{h}_I(f_2, t)\tilde{h}_J(f_2, t) \Big\rangle \Big)\Bigg\rangle,
\end{aligned}
\end{equation}
where $\tilde{h}_I(f)$ is the Fourier transform of strain $h_I(t)$, and $\mathcal{\hat{C}}_{S}^{IJ}(f_1,f_2)$ is the measured value of the spectral covariance between frequencies $f_1$ and $f_2$. In the noise-dominated regime, the covariance of $\hat{\mathcal{C}}_{S}^{IJ}(f_1,f_2)$ is given by 
\begin{equation}
    \begin{aligned}
        \Big(\hat{\Sigma}^{IJ}_{t}(f_1,f_2)\Big)^{2} =& \Big\langle \Big( \hat{\mathcal{C}}_{S,t}^{IJ}(f_1,f_2) - \Big\langle \hat{\mathcal{C}}_{S,t}^{IJ}(f_1,f_2) \Big\rangle\Big)^{2}\Big\rangle\\
        \equiv & ~\Big\langle \Big(\tilde{n_{I}}(f_1) \tilde{n_{J}}(f_1) \tilde{n_{I}}(f_2) \tilde{n_{J}}(f_2)\Big)^{2}\Big\rangle\\
        = & ~\mathcal{N}_{I}(f_1) \mathcal{N}_{J}(f_1) \mathcal{N}_{I}(f_2)  \mathcal{N}_{J}(f_2), 
    \end{aligned}
\end{equation}
where $\tilde{n}_{I}(f_1)$ is the Fourier transform of the time-domain noise, $n_I(t)$. $\mathcal{N}_{I}(f_1)$ is the noise power spectrum of the detector I.

\newpage

\end{document}